\documentclass[aps,prb,twocolumn,showpacs,showkeys,preprintnumbers]{revtex4}
\usepackage[dvips]{graphicx}
\usepackage{amssymb}
\usepackage{amsmath}
\usepackage{enumerate}
\usepackage{color}

\newcommand{\ee}{\end{equation}}
\newcommand{\be}{\begin{equation}}
\newcommand{\bea}{\begin{eqnarray}}
\newcommand{\eea}{\end{eqnarray}}

\newcommand{\tx}{\tilde{x}}

\newcommand{\eu}{{\rm e}}
\newcommand{\ii}{{\rm i}}

\newcommand{\sn}{{\rm sn}}
\newcommand{\cn}{{\rm cn}}
\newcommand{\dn}{{\rm dn}}
\newcommand{\Tr}{{\rm Tr}}
\newcommand{\Ord}{{\mathcal O}}

\begin{document}

\preprint{MIT-CTP 4349}

\title{Correlation length and  unusual corrections to the entanglement entropy}

\author{Elisa Ercolessi$^{1}$}
\email{ercolessi@bo.infn.it}

\author{Stefano Evangelisti$^{1,2}$}
\email{stefano.evangelisti@worc.ox.ac.uk}

 \author{Fabio Franchini$^{3,4}$}
\email{fabiof@mit.edu}

\author{Francesco Ravanini$^{1}$}
\email{ravanini@bo.infn.it}
\affiliation{ }

\affiliation{$^{1}$ Department of Physics, University of Bologna and I.N.F.N. Sezione di Bologna,
 Via Irnerio 46, 40126, Bologna, Italy}
\affiliation{$^{2}$ The Rudolf Peierls Centre for Theoretical Physics, Oxford University, Oxford, UK }
\affiliation{$^{3}$ Department of Physics, Massachusetts Institute of Technology, Cambridge, MA 02139, U.S.A. }
\affiliation{$^{4}$ SISSA and I.N.F.N, Via Bonomea 265, 34136, Trieste, Italy}

\begin{abstract}
We study analytically the corrections to the leading terms in the R\'enyi entropy of a massive lattice theory, showing significant deviations from na\"ive expectations. In particular, we show that finite size and finite mass effects give rise to different contributions (with different exponents) and thus violate a simple scaling argument.

In the specific, we look at the entanglement entropy of a bipartite {\em XYZ} spin-$1/2$  chain in its ground state. When the system is divided into two semi-infinite half-chains, we have an analytical expression of the R\'enyi entropy as a function of a single mass parameter. In the scaling limit, we show that the entropy as a function of the correlation length formally coincides with that of a bulk Ising model. This should be compared with the fact that, at criticality, the model is described by a $c=1$ conformal field theory and the corrections to the entropy due to  finite size effects show exponents depending on the compactification radius of the theory. We will argue that there is no contradiction between these statements.

If the lattice spacing is retained finite, the relation between the mass parameter and the correlation length generates new subleading terms in the entropy, whose form is path-dependent in phase-space and whose interpretation within a field theory is not available yet. These contributions arise as a consequence of the existence of stable bound states and are thus a distinctive feature of truly interacting theories, such as the {\em XYZ} chain.
\end{abstract}

\keywords{Entanglement in extended quantum systems, Integrable spin chains, Integrable quantum field theory}

\pacs{03.65.Ud, 02.30.Ik, 75.10.Pq, 11.10.-z}

\maketitle

\section{Introduction}

In the past few years, there has been a growing interest in quantifying the degree of ``quantum-ness'' of a many-body state, which is usually taken as the ground state $|0\rangle$ of a given Hamiltonian \cite{Fazio, eisert, ccdrev}. As the entanglement constitutes an intrinsically quantum property, one popular way to measure this is through  {\it bipartite entanglement entropy} \cite{Schumacher, nielsen}. To this end, the system is divided into two subsystems ($A$ and $B$) and one looks at the reduced density matrix, obtained by tracing out one of the two subsystems
\be
   \hat{\rho}_A \equiv \Tr_B |0 \rangle \langle 0| \; .
\ee
We will be interested in the R\'enyi
	 entropies  \cite{Renyi}
\be
   {\cal S}_{\alpha} \equiv {1 \over 1-\alpha} \ln \Tr \hat{\rho}_A^\alpha \; ,
   \label{eq:renyi1}
\ee
which, in the $\alpha \to 1$ limit, reduces to the Von Neumann entropy
\be
   {\cal S} \equiv - \Tr \hat{\rho}_A \log \hat{\rho}_A \; .
\ee
Varying the parameter $\alpha$ in (\ref{eq:renyi1}) gives us access to a lot of information on $\hat{\rho}_A$, including its full spectrum \cite{calabreselefevre2008, franchini2010}.

For gapped systems, the entanglement entropy satisfies the so-called {\it area law}, which means that its leading contribution for sufficiently large subsystems is proportional to the area of the boundary separating system $A$ from $B$. In $1+1$ dimensional systems, the area law implies that the entropy asymptotically saturates to a constant (the boundary between regions being made just by isolated points).

Critical systems can present deviations from the simple area law. In one dimension, in particular, the entanglement entropy of systems in the universality class of a conformal field theory (CFT) is known to diverge logarithmically with the subsystem size \cite{Holzey, Cardy}. From CFT, a lot is known also about the subleading corrections, which, in general, take the {\em unusual} form \cite{calabrese2010, camplostrini2010, cardy2010, essler2010}
\be
   {\cal S}_\alpha (\ell) = {c + \bar{c} \over 12} \left( 1 + {1 \over \alpha} \right) \ln {\ell \over a_0} + c'_\alpha
   + b_\alpha (\ell) \,\ell^{-2 h/\alpha} + \ldots \; ,
   \label{Sncriticalexp}
\ee
where $c$ is the central charge of the CFT, $\ell$ is the length of subsystem $A$, $a_0$ is a short distance cutoff, $c'_\alpha$ and $b_\alpha (\ell)$ are non-universal and the latter includes a periodic function of $\ell$ (with the period given by the Fermi momentum \cite{camplostrini2010, essler2010}), and $h$ is the scaling dimension of the operator responsible for the correction (relevant or irrelevant, but not marginal, since these operators generate a different kind of correction, which will be discussed later). This result is achieved using replicas, and thus, strictly speaking, requires $\alpha$ to be an integer. Moreover, it should be noted that in Ref.\cite{cardy2010} the corrections are obtained from dimensionality arguments, by regularizing divergent correlations by an ultra-violet cut-off $a_0$. Thus, technically, the subleading contributions in (\ref{Sncriticalexp}) are extracted from scaling properties and are all of the form $\ell/a_0$.

Determining the exponents of the corrections is important both in fitting numerics (where often really large $\ell$ are unobtainable) and also for a better understanding of the model. For instance, the scaling exponent $h$ also determines the large $n$ limit of the entropy ({\it single copy entanglement}) \cite{calabrese2010}. Moreover, especially for $c=1$ theories, $h$ provides a measure of the compactification radius of the theory\cite{essler2010} and thus of the decaying of the correlation functions. Up to now, this conjecture has been checked in a variety of critical quantum spin chains models \cite{fagotti,alcaraz2011,dalmonte2011}.

Moving away from a conformal point, in the gapped phase universality still holds for sufficiently small relevant perturbations. Simple scaling arguments guarantee that the leading terms survive, but with the correlation length $\xi$ replacing the infra-red length-scale $\ell$. Recent results, based on exactly solvable models, indicate the appearance of the same kind of unusual corrections to the R\'enyi
	 entropies, which are now functions of the correlation length $\xi$ with the same exponent $h$ \cite{calabrese2010}:
\be
   {\cal S}_\alpha = {c \over 12} \left( {1 + \alpha \over \alpha} \right) \ln {\xi \over a_0} + A_\alpha
   + B_\alpha \xi^{-h/\alpha} + \ldots \; .
   \label{Sexp}
\ee
There is a factor of $2$ difference in each term between (\ref{Sncriticalexp}) and (\ref{Sexp}), due to the fact that the first is a bulk theory (with both chiralities in the CFT), while the latter is expected to be akin to a boundary theory, were only one chirality in the light-cone modes effectively survives.

In this paper, we are going to investigate the subleading terms in the R\'enyi entropy of the one-dimensional {\em XYZ} model. In the scaling limit we find that the entropy is (modulo a multiplicative redefinition of the correlation length)
\bea
    \hskip-2.0cm {\cal S}_\alpha & = & {1 + \alpha \over 12 \alpha} \ln {\xi \over a_0} - {1 \over 2} \ln 2
    \label{Sxiexp}\\
    && -{1 \over 1 - \alpha} \sum_{n=1}^\infty \sigma_{-1} (n) \left[ \left( {\xi \over a_0} \right)^{-{2n \over \alpha}}
    - \left( {\xi \over a_0} \right)^{-{4n \over \alpha}} \right]
    \nonumber \\
    &&  + {\alpha \over 1 - \alpha} \sum_{n=1}^\infty \sigma_{-1} (n) \left[ \left( {\xi \over a_0} \right)^{-2n}
    - \left( {\xi \over a_0} \right)^{-4n} \right] \: ,
    \nonumber
\eea
where $\sigma_{-1} (n)$ is a {\it divisor function}, defined by (\ref{wndef}).

While the leading term correctly reproduces a $c=1$ central charge, the interpretation of the scaling exponents in the subleading addenda is less straightforward. Comparing the exponent of the first correction with (\ref{Sexp}) would indicate $h=2$, i.e. a marginal operator. This term cannot be due to the same operator acting at the critical point since, as shown in Ref. \cite{cardy2010}, marginal operators give rise to logarithmic corrections. Moreover, the critical {\em XXZ} chain is known to have relevant fields with $h=K$ (the Luttinger parameter) and the opening of a gap in the {\em XYZ} model implies the presence of a $h <2$ operator.
We will show that the operator content that can be extracted from (\ref{Sxiexp}) matches  that of a {\em bulk Ising} model  and the first correction can be interpreted as arising from the energy field. Notice that the leading, logarithmic term can thus be equally interpreted as $ {c + \bar{c} \over 12}$ with $c = \bar{c} = {1 \over 2}$.

	Furthermore, if we include also lattice effects, which vanish in the strict scaling limit, additional corrections appear in (\ref{Sxiexp}) and, while they are less important than the dominant one for sufficiently large $\alpha$, they can be relevant for numerical simulations in certain ranges of $\alpha$. These corrections turn out to be path dependent (probably due to the action of different operators) and many kind of terms can arise, such as $\xi^{-2h}$, $\xi^{-2h/\alpha}$, $\xi^{-(2-h)}$, $\xi^{-2(1+1/\alpha)}$ or even $1/\ln(\xi)$. In light of Ref.~\cite{cardy2010}, some of these terms were to be expected, but the others still lack a field theoretical interpretation, which might be possible by applying a reasoning similar to that of Ref.~\cite{cardy2010} for a sine-Gordon model.

The paper is organized as follows: in Sec. \ref{sec:XYZ} we introduce the {\em XYZ} model, its phase-diagram and elementary excitations. In Sec. \ref{sec:rho} we discuss the structure of the reduced density matrix obtained by tracing out of the ground state half of the system and its formal equivalence with the characters of an Ising model. In Sec. \ref{sec:exp} we present the full expansion of the R\'enyi
	 entropy and discuss how in the scaling limit this expansion coincides with that of a bulk Ising model. Then we discuss the lattice corrections both in the regime where the low-energy excitations are free particles and where they are bound states. In the latter case, we show that different paths of approach to the critical point give rise to different sub-subleading corrections. Finally, in Sec. \ref{sec:end} we discuss our results and their meaning.

\section{The {\em XYZ} spin chain}
\label{sec:XYZ}

We will consider the quantum spin-$\frac{1}{2}$ ferromagnetic {\em XYZ} chain, which is described by the following Hamiltonian
\begin{equation}
   \hat{H}_{XYZ} =
   - J {\displaystyle \sum_{n}} \left[ \sigma_{n}^{x} \sigma_{n+1}^{x}
   + J_{y} \sigma_{n}^{y} \sigma_{n+1}^{y} + J_{z} \sigma_{n}^{z} \sigma_{n+1}^{z} \right] \; ,
   \label{eq:XYZ1}
\end{equation}
where the $\sigma_{n}^{\alpha}$ ($\alpha=x,y,z$) are the Pauli matrices acting on the site $n$ and the sum ranges over all sites $n$ of the chain;  the constant $J$, which we take to be positive, is an overall energy scale while  $(J_y, J_z)$ take into account the degree of anisotropy of the model.

\begin{figure}
   \begin{centering}
   \includegraphics[width=8cm]{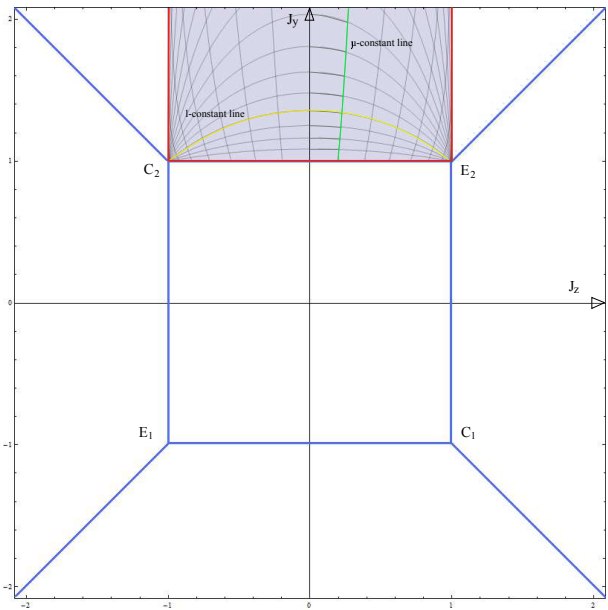}
   \end{centering}
   \caption{(Colored online) Phase Diagram of the {\em XYZ} model in the $(J_z, J_y)$ plane. The blue solid lines --$J_z = \pm 1$, $|J_y| \le 1$; $J_y= \pm 1$, $|J_z| \le 1$ and $J_z = \pm J_y$, $|J_z| \ge 1$-- correspond to the critical phase of a rotated {\em XXZ} chain. Out of the four ``tricritical'' points, $C_{1,2}$ are conformal and $E_{1,2}$ are not. The area within the red rectangle is the portion of the phase diagram we study in this article, and it is completely equivalent to the principal regime of the eight-vertex model. The yellow and green lines are the curves of constant $l$ and $\mu$ respectively in the $(J_z,J_y)$ plane, according to the parametrization (\ref{ellipticparam}).}
   \label{fig:phasediagram}
\end{figure}

In Fig.~\ref{fig:phasediagram} we draw a cartoon of the phase diagram of the {\em XYZ} chain. The model is symmetric under reflections along the diagonals in the $(J_z, J_y)$ plane. The system is gapped in the whole plane, except for six critical half-lines/segments: $J_z = \pm 1$, $|J_y| \le 1$; $J_y= \pm 1$, $|J_z| \le 1$ and $J_z = \pm J_y$, $|J_z| \ge 1$. All of these lines correspond to the paramagnetic phase of an $XXZ$ chain, but with the anisotropy along different directions. Thus, in the scaling limit they are described by a $c=1$ CFT, with compactification radius varying along the line. We will use $0 \le \beta \le \sqrt{8 \pi}$ (the sine-Gordon parameter of the corresponding massive theory) to parametrize the radius. The critical segments meet three by three at four ``tricritical'' points. At these endpoints, $\beta$ is the same along each line, but the different phases have a rotated order parameter. Two of these points --$C_{1,2}=(1,-1),(-1,1)$-- are conformal points with $\beta^2 = 8 \pi$; while the other two --$E_{1,2}=(1,1),(-1,-1)$-- correspond to $\beta^{2} = 0$ and are {\it nonconformal}. The former points correspond to an antiferromagnetic (AFM) Heisenberg chain at the BKT transition, while the latter describe the Heisenberg ferromagnet. Thus, at $E_{1,2}$ the system undergoes a transition in which the ground state passes from a disordered state to a fully aligned one. Exactly at the transition, the ground state is highly degenerate while the low-energy excitations are gapless magnons with a quadratic dispersion relation.

In studying the {\em XYZ} chain, one takes advantage of the fact that (\ref{eq:XYZ1}) commutes with the transfer matrices of the the zero-field eight-vertex model (see for example Refs. \cite{Sutherland, Baxter}) and thus the two systems can be solved simultaneously. The solution of the latter is achieved through the parametrization of $J_y, J_z$ in terms of elliptic functions
\begin{equation}\begin{array}{lcl}
   J_y & = & -\Delta \equiv \frac{\mbox{cn}(\ii \lambda)\;\mbox{dn}(\ii \lambda)}{1-k\;\mbox{sn}^{2}(\ii \lambda)} \; \\
   J_z & = & -\Gamma \equiv -\frac{1+k\;\mbox{sn}^{2}(\ii \lambda)}{1-k\;\mbox{sn}^{2}(\ii \lambda)} \; ,
   \label{eq:XYZ3bis}
	\end{array}
\end{equation}
where $(\Gamma,\Delta)$ are the well-known Baxter parameters \cite{Baxter}, and $\mbox{sn}(x)$, $\mbox{cn}(x)$ and $\mbox{dn}(x)$ are Jacobian elliptic functions of parameter $k$.
$\lambda$ and $k$ are parameters, whose natural domains are
\begin{equation}
   0<k<1 \; , \qquad 0 \le \lambda \le I(k') \; ,
   \label{eq:XYZ4}
\end{equation}
$I(k')$ being the complete elliptic integral of the first kind of argument $k' \equiv \sqrt{1-k^{2}}$.

The definition of $(\Delta,\Gamma)$ itself is particularly suitable to describe the anti-ferroelectric phase of the eight-vertex model (also referred to as the {\it principal regime}), corresponding to $\Delta\le-1$ and $|\Gamma|\le1$. However, using the symmetries of the model and the freedom under the rearrangement of parameters, it can be applied to the whole of the phase diagram of the spin Hamiltonian (for more details see Ref.~\cite{Baxter}). For the sake of simplicity, in this paper we will focus only on the {\em rotated} principal regime: $J_y \ge 1$, $|J_z|\le 1$, see Fig.~\ref{fig:phasediagram}, and we defer to a different publication some interesting properties of the generalization to the whole phase diagram.

Before we proceed, it is more convenient to switch to an elliptic para\-me\-trization equivalent to (\ref{eq:XYZ3bis}):
\be
   l \equiv {2 \sqrt{k} \over 1 + k} \; , \qquad
   \mu \equiv \pi {\lambda \over I(k')} \; .
\ee
The elliptic parameter $l$ corresponds to a gnome $\tau \equiv \ii {I(l') \over I(l)} = \ii {I(k') \over 2 I(k)}$, which is half of the original. The relation between $k$ and $l$ is known as {\it Landen transformation}. Note that $0 \le \mu \le \pi$.

In terms of these new parameters, we have
\be
   \Gamma = {1 \over \dn [ 2 \ii I(l') \mu / \pi  ; l ] } \; , \qquad
   \Delta = - { \cn [2 \ii I(l') \mu / \pi ; l] \over \dn [2 \ii I(l') \mu / \pi ; l] } \; .
   \label{ellipticparam}
\ee

Curves of constant $l$ always run from the AFM Heisenberg point at  $\mu=0$ to the isotropic ferromagnetic point at $\mu= \pi$. For $l=1$ the curve coincides with one of the critical lines discussed above, while for $l=0$ the curve run away from the critical one to infinity and then back. In Fig.~\ref{fig:phasediagram} we draw these curves for some values of the parameters.

For later convenience, we also introduce
\be
  x \equiv \exp \left[ - \pi {\lambda \over 2 I (k)} \right] = \eu^{\ii \mu \tau} .
  \label{qxdef}
\ee
Using Jacobi's theta functions, we introduce the elliptic parameter $k_1$ connected to $x$, i.e.
\be
  k_1 \equiv {\theta_2^2 (0, x) \over \theta_3^2 (0,x)}
  =  {x^{1\over 2} \over 4} {(-1;x^2)^4_\infty \over (-x;x^2)^4_\infty}
  \equiv k (x)  \; ,
  \label{k1def}
\ee
or, equivalently, $\pi {I(k'_1) \over I(k_1)} = - \ii \mu \tau$ (i.e., $l$ is to $\tau$ what $k_1$ is to $\mu \tau / \pi$). In (\ref{k1def}) we also used the {\em q}-Pochhammer symbol
\be
   (a;q)_n \equiv \prod_{k=0}^{n-1} (1 - a q^k) \; .
\ee

The correlation length and the low-energy excitations of the {\em XYZ} chain were calculated in Ref. \cite{McCoy}.
There are two types of excitations. The first can be characterized as free quasi-particles (spinons). The lowest band is a 2-parameter continuum with
\bea
  \Delta E_{\rm free} (q_1,q_2) & = & - J \: {\sn [2 I(l') \mu /\pi;l'] \over I(l) } \; I(k_1) \times \\
  \label{scaling7}
  && \hskip -1cm \left( \sqrt{1 - k_1^2 \cos^2 q_1} + \sqrt{1 - k_1^2 \cos^2 q_2} \right) \; .
  \nonumber
\eea
The energy minimum of these state is achieved for $q_{1,2}=0, \pm \pi$ and gives a mass gap
\bea
  \hskip -0.6cm \Delta E_{\rm free} & = & 2 J  {1 \over I(l)} \sn \left[ 2 I(l') {\mu  \over \pi} ; l' \right] \; I(k_1) k'_1 \; .
  \label{freegap}
\eea

For $\mu > \pi/2$, in addition to the free states just discussed, some bound states become progressively stable.
They are characterized by the following dispersion relation
\bea
  \label{scaling11}
  \Delta E_s (q) & = & - 2 J
  {\sn [2 I(l') \mu /\pi;l'] \over I(l) } \:
  { I(k_1) \over \sn(s y;k'_1)}
  \nonumber \\
  & \times & \sqrt{1 - \dn^2 (s y;k_{1}') \, \cos^2 {q \over 2}}
  \nonumber \\
  & \times & \sqrt{1 - \cn^2 (s y;k_{1}') \, \cos^2 {q \over 2}} \; ,
\eea
where $y \equiv \ii I(k_1) \tau \left( {\mu \over \pi} - 1 \right)$ and $s$ counts the number of quasi-momenta in the string state. In the scaling limit, these bound states become breathers.
The mass-gap for the bound states is (setting $q=0$ above)
\bea
  \Delta E_s & = & \Delta E_{\rm free} \: \sn (s y;k'_1) \; ,
  \label{boundgap}
\eea
from which one sees that for $\mu > \pi/2$ the $s=1$ bound state becomes the lightest excitation.

\begin{figure}
   \includegraphics[width=\columnwidth]{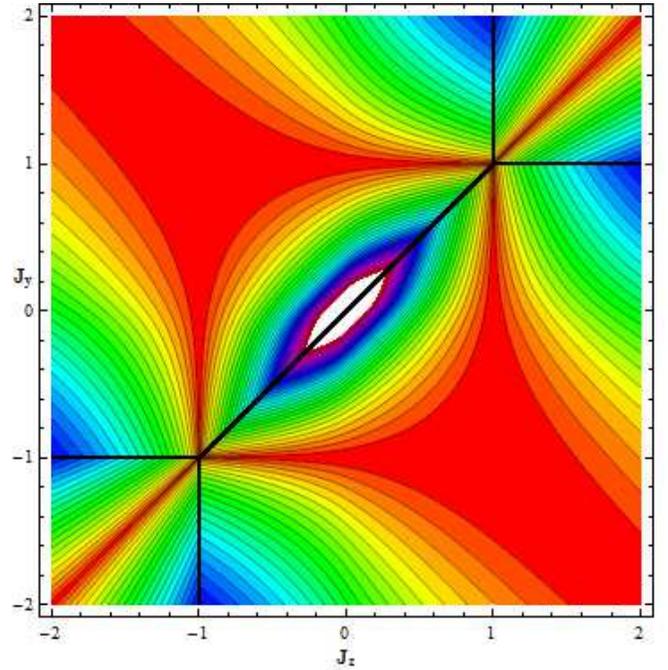}
   \caption{Curves of constant correlation length of the {\em XYZ} model in the ($J_z,J_y$) plane. Regions of similar colors correspond to the same  correlation length values and red colors are associated to higher values of $\xi$. It is worth stressing that the correlation length does not show any essential critical behavior, whilst the block entropy shows it in proximity of points E$_{1}$ and E$_{2}$.}
   \label{fig:correlation_length}
\end{figure}

The correlation length for the {\em XYZ} chain (Fig.~\ref{fig:correlation_length}) was also calculated in Ref.~\cite{McCoy} and it is given by:
\be
   \xi^{-1} = {1 \over a_0} \left\{ \begin{array}{ll}
        -\frac{1}{2}\ln k_{2}
        & 0 \leq \mu \leq \frac{\pi}{2} \; ,\cr
        -\frac{1}{2} \ln \frac{k_{2}} {\dn^{2} \left[ \ii 2 I(k_{2}) {\tau \over \pi} \left( \mu  - { \pi \over 2} \right) ; k'_{2} \right]}
        & \frac{\pi}{2} < \mu \leq \pi \; , \cr
   \end{array} \right.
   \label{xibound}
\ee
where $a_0$ is a short distance cut-off, such as the lattice spacing, that sets the length unit and the new parameter $k_2$ is the Landen transformed of $k_1$:
\be
  k_2 \equiv k (x^2) = {1 - k'_1 \over 1 + k'_1} \; .
  \label{k2def}
\ee

The first behavior in (\ref{xibound}) is due to the free particles states, while the $s=1$ bound state is responsible for the second \cite{comment}.

\section{The Reduced Density Matrix}
\label{sec:rho}

The bipartite entanglement entropy for the ground state of the {\em XYZ} chain was calculated in \cite{Evangelisti, Ercolessi2011}, in the limit where the infinite chain is partitioned in two (semi-infinite) half lines. For this configuration, the reduced density matrix can be computed as the product of the four corner transfer matrices (CTM) of the corresponding eight-vertex model \cite{Peschel1, Nishino1, Nishino2}. In Ref. \cite{Evangelisti} it was shown that it can be written as
\be
   \hat{\rho} = {1 \over {\cal Z}} \bigotimes_{j=1}^\infty
                \left( \begin{array}{cc}
                    1 & 0 \\
                    0 & x^{2j}
                \end{array} \right) \; ,
   \label{XYZrho}
\ee
where $\mathcal{Z} \equiv (-x^2,x^2)_\infty$, that is, the partition function of the eight-vertex model, is the normalization factor that ensures that $\Tr \hat{\rho} =1$.
Thus we have
\be
   \label{eq:renyi3}
   \mbox{Tr} \hat{\rho}^{\alpha} = { \left( -x^{2 \alpha}; x^{2 \alpha} \right)_\infty \over
   \left( -x^2;x^2 \right)^\alpha_\infty}
\ee
and for the R\'enyi
	 entropy
\be
   {\cal S}_{\alpha} =
   \frac{\alpha}{\alpha-1} \sum_{j=1}^{\infty} \ln \left ( 1 + x^{2 j} \right)
   + \frac{1}{1-\alpha} \sum_{j=1}^{\infty} \ln \left ( 1 + x^{2 j \alpha} \right) \; .
   \label{eq:renyi4}
\ee

The structure (Eqs. \ref{XYZrho}, \ref{eq:renyi4}) for the reduced density matrix of the half-line is common to all integrable, local spin-$1/2$ chains \cite{Baxter} and thus the entanglement spectrum of these models is the same and only depends on $x$, which in this context is usually parametrized as $x = \eu^{-\epsilon}$. For the {\em XYZ} chain, $\epsilon = - \ii \mu \tau$. In Ref. \cite{Ercolessi2011} we showed that $\epsilon$ (and thus the entropy) has an essential singularity at non-conformal points, and thus its behavior differs dramatically from the conformal one.

\begin{figure}
   \includegraphics[width=\columnwidth]{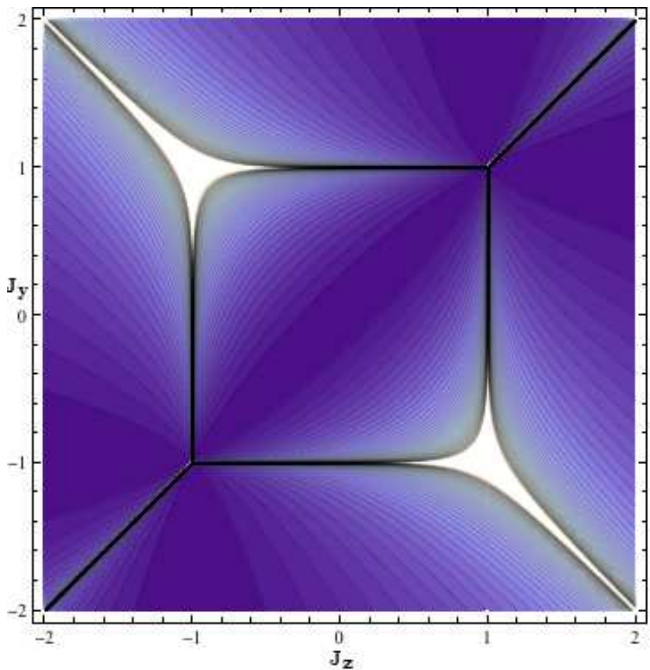}
   \caption{Curves of constant entropy of the {\em XYZ} model in the ($J_z,J_y$) plane. Regions of similar colors have similar entropy values and the line where colors change are the lines of constant entropy. The brighter is the color, the bigger the entropy.}
   \label{fig:isoentropy}
\end{figure}

From (\ref{eq:renyi4}), one can see that the R\'enyi
	 entropy is a monotonically decreasing function of $\epsilon$:
\be
   \lim_{\epsilon \to 0} {\cal S}_\alpha = \infty \; , \qquad \qquad
   \lim_{\epsilon \to \infty} {\cal S}_\alpha = 0 \; .
\ee
Using (\ref{qxdef}) and \cite{Baxter}
\be
   \label{ste8}
   l = \sqrt{ {1-\Gamma^{2} \over \Delta^{2}-\Gamma^{2} } } \;, \qquad
   \dn \left[ 2 \ii I(l') {\mu \over \pi} ;l \right] = {1\over \Gamma}
\ee
we can plot the entanglement entropy in the phase diagram of the {\em XYZ} model. In Fig \ref{fig:isoentropy} we show a contour plot of the Von Neumann entropy in the $(J_z,J_y)$ plane, from which one can clearly see the different behavior of the conformal and non-conformal points.

{\em q}-products of the form (\ref{eq:renyi3}) give easy access to the spectral distribution of the reduced density matrix \cite{franchini2010}, since
\be
   (-q,q)_\infty = \prod_{k=1}^\infty (1 + q^k) = 1 + \sum_{n=1}^\infty p^{(1)} (n) q^n \; ,
   \label{partition1}
\ee
where $p^{(1)} (n)$ is the number of partitions of $n$ in distinct positive integers. Also note that, since
\be
  \prod_{k=1}^\infty (1 + q^k) = \prod_{k=1}^\infty (1 - q^{2k -1})^{-1} \; ,
  \label{prodrel}
\ee
$p^{(1)} (n) = p_{\cal O} (n)$, that is,  the number of partitions of $n$ into positive {\it odd} integers.

Moreover, one can recognize ${\cal Z}$ to be formally equal to the character of the spin field of the Ising CFT.
To show this, we write
\be
   {\cal Z} \left( q=x^2 \right)= \prod_{j=1}^\infty (1 + q^j)
   = \prod_{j=1}^\infty {1 - q^{2j} \over 1 - q^j}
\ee
and we use the Euler's formula for pentagonal numbers (which is a consequence of the Jacobi triple-product Identity) \cite{Wittaker}
\bea
  \prod_{j=1}^\infty \left( 1 - q^{2j} \right)
  & = & \sum_{n=-\infty}^\infty (-1)^n q^{n(3n-1)}
  \label{pentagonal}
  \\
  & = & \sum_{n=-\infty}^\infty \left[ q^{2n(6n-1)} - q^{(2n+1)(6n+2)} \right]
   \nonumber
\eea
to recognize that
\be
   {\cal Z} = x^{-{1 \over 12}} \chi_{1,2}^{\rm Ising} \left( \ii \epsilon / \pi \right) \; ,
   \label{ZchiIsing}
\ee
where $ \chi_{1,2}^{\rm Ising} (\tau)$ is the character of the spin $h_{1,2} = 1/16$ operator of a $c=1/2$ CFT:
\bea
   \hskip-0.5cm \chi_{r,s}^{(p,p')} (\tau) & = &
   {q^{-1/24} \over \prod_{j=1}^\infty (1 - q^j)} \sum_ {n=-\infty}^\infty \left[
   q^{[2 p p' n + r p - s p']^2 \over 4 p p'} \right.
   \nonumber \\
   && \qquad \qquad \qquad \left.
   -  q^{[2 p p' n + r p + s p']^2 \over 4 p p'} \right] \; ,
   \label{characterdef}
\eea
with $q \equiv \eu^{2 \ii \pi \tau}$ and $(p,p')=(4,3)$ for the Ising minimal model.
We have
\be
   \mbox{Tr} \hat{\rho}^{\alpha} = { \chi_{1,2}^{\rm Ising} (\ii \alpha \epsilon / \pi) \over
   \left[ \chi_{1,2}^{\rm Ising} (\ii \epsilon / \pi) \right]^\alpha  }  \: .
   \label{isingrho}
\ee

As we discussed above, the critical line of the {\em XXZ} chain is approached for $l \to 1$, that is, for $\tau \to 0$ and $x \to 1$.
On this line, excitations are gapless and in the scaling limit the theory can be described by a conformal field theory with central charge $c=1$. To each $0<\mu<\pi$ it correspond a different point on the critical line, with sine-Gordon parameter $\beta^2 = 8 \pi \left( 1 - {\mu \over \pi} \right)$ \cite{Luther}. The two endpoints are exceptions, since $\mu=0,\pi$ identify the same points for every $l$. However, while in the conformal one we still have $x \to 1$, close to the ferromagnetic point, around $\mu = \pi$, both $x$ and $q$ can take any value between $0$ and $1$. Hence a very different behavior of the entanglement entropy follows.\cite{Ercolessi2011, Franchini}

In order to study the asymptotic behavior of the entanglement entropy close to the conformal points, it is convenient to use the dual variable
\be
   \tx \equiv \eu^{- \ii {\pi^2 \over \mu \tau}} = \eu^{- {\pi^2 \over \epsilon}} \; .
   \label{qxprimedef}
\ee
which is such that  $\tx \to 0$ as $l \to 1$.

Expressions like (\ref{eq:renyi3}) involving q-products can be written in terms of elliptic theta functions. For instance, this was done for the entanglement entropy of the {\em XYZ} in \cite{Ercolessi2011, Franchini2}. To study the conformal limit, one performs a modular transformation that switches $x \to \tx$.
Since
\be
   k (\tx) = k' (x) = k'_1 = { (x;x^2)^4_\infty \over (-x;x^2)^4_\infty } \; ,
\ee
and using (\ref{k1def}), we have
\bea
   \left( -x^{2 \alpha}; x^{2 \alpha} \right)_\infty & = &
   \left[ {k^2 \left( x^\alpha \right) \over 16 x^\alpha k' \left( x^\alpha \right) } \right]^{1/12}
   \nonumber \\
   = \left[ { k^{\prime 2} \left( \tx^{1/\alpha } \right)
   \over 16 x^\alpha k \left( \tx^{1/\alpha} \right) } \right]^{1/12}
   & = & { \left( \tx^{1/\alpha} ;  \tx^{2/\alpha} \right)_\infty \over
   2^{1/2} x^{\alpha / 12} \tx^{1/24\alpha} } \; .
\eea
Thus
\be
   \Tr \hat{\rho}^\alpha = 2^{{\alpha -1 \over 2}} \tx^{\alpha^2 -1 \over 24\alpha}
   { \left( \tx^{1/\alpha};  \tx^{2/\alpha} \right)_\infty \over
   \left( \tx; \tx^2 \right)^\alpha_\infty } \; .
   \label{dualrho}
\ee

The modular transformation that allowed us to switch from $x$ to $\tx$ is the same one that connects characters in minimal model of inverse temperature. For the spin operator of the Ising model we have
\be
   \chi_{1,2}^{\rm Ising} (\tau) = {1 \over \sqrt{2}} \left[  \chi_{1,1}^{\rm Ising} (-1/ \tau)
   - \chi_{2,1}^{\rm Ising} (-1 / \tau) \right] \; .
\ee
Using (\ref{characterdef}) and the identities (\ref{prodrel} and  \ref{pentagonal}) one can prove that
\be
   \chi_{1,2}^{\rm Ising} \left( \ii \epsilon / \pi \right) = {1 \over \sqrt{2}} \tx^{-{1 \over 24}} \left( \tx; \tx^2 \right)_\infty \; ,
   \label{dualqprodZ}
\ee
which agrees with (\ref{dualrho}) and implies
\be
    \Tr \hat{\rho}^\alpha = 2^{\alpha -1 \over 2} { \chi_{1,1}^{\rm Ising} (\ii \pi / \alpha \epsilon)
    -  \chi_{2,1}^{\rm Ising} (\ii \pi / \alpha \epsilon)
    \over \left[  \chi_{1,1}^{\rm Ising} (\ii \pi / \epsilon)
    -  \chi_{2,1}^{\rm Ising} (\ii \pi / \epsilon) \right]^\alpha } \; .
    \label{dualrhochar}
\ee
This agrees with what conjectured in Ref.\cite{calabrese2010}, but with the important difference that the characters in (\ref{dualrhochar}) are $c=1/2$ and do not belong to the infrared $c=1$ bulk description of the {\em XYZ} chain \cite{Saleur}. This Ising character structure for the CTM of the eight-vertex model was already noticed, see Ref. \cite{cardyle}. We acknowledge that, once the formal equivalence (\ref{ZchiIsing}) is understood, the content of the last page follows almost trivially, but  we decided to provide a brief derivation here for the sake of completeness and because this result has consequences on the structure of the entanglement entropy, a fact which is not well known and will be discussed in the next paragraph. Finally, we notice that, being only a formal equivalence, eq. (\ref{ZchiIsing}) does not imply any underlying Virasoro algebra at work for CTM (as far as we know) and it is thus important to recognize that these manipulations stand on more general mathematical concepts.

\section{Expansions of the Entanglement Entropy} \label{se:expansion}
\label{sec:exp}

Close to the conformal points, Eq. (\ref{eq:renyi4}) is just a formal series, since $x \simeq 1$. However, using (\ref{dualrho}), it is straightforward to write a series expansion for the R\'enyi
	 entropy (\ref{eq:renyi1}) in powers of $\tx \ll 1$:
\bea
    \hskip-2.0cm {\cal S}_\alpha & = & - {1 + \alpha \over 24 \alpha} \ln \tx - {1 \over 2} \ln 2
    \label{Stxexp}\\
    && -{1 \over 1 - \alpha} \sum_{n=1}^\infty \sigma_{-1} (n) \left[ \tx^{n \over \alpha}
    - \alpha \tx^n - \tx^{2n \over \alpha} + \alpha \tx^{2n} \right] \: ,
    \nonumber
\eea
where the coefficients
\be
   \sigma_{-1} (n) \equiv {1 \over n} \sum_{\substack{ j < k=1 \\ j \cdot k =n}}^\infty (j + k)
   + \sum_{\substack{ j=1 \\ j^2 =n}}^\infty {1 \over j} = {\sigma_1 (n) \over n}
   \label{wndef}
\ee
is a divisor function \cite{apostolbook} and takes into account the expansion of the logarithm over a q-product and play a role similar to the partitions of integers in (\ref{partition1}).
It is worth noticing that the constant term $\ln(2^{-1/2})\equiv \ln(S_{1/16}^{0})$ - where $S_{1/16}^{0}$ is an element of the modular S-matrix of the Ising model - is the contribution to the entropy due to the boundary \cite{Affleck}.

The $\alpha \to 1$ yields the Von Neumann entropy:
\bea
    \hskip-2.0cm {\cal S}_\alpha & = & - {1 \over 12} \ln \tx - {1 \over 2} \ln 2
    \label{SVNtxexp}\\
    && - \sum_{n=1}^\infty \sigma_{-1} (n) \left[ n \left(  \tx^n - 2 \tx^{2n} \right) \ln \tx
    + \tx^{n} - \tx^{2n} \right] \: .
    \nonumber
\eea
We see that, contrary to what happens for $\alpha>1$, all subleading  terms -which are powers of $\tx$- acquire a logarithmic correction, which strictly vanishes only at the critical points. A more specific analysis of this phenomenon, together with a quantitative study of what introduced in the next subsections, will be the subject of a future, partly numerical, publication.

\subsection{Scaling limit}

Comparing with (\ref{Sexp}) and coherently with (\ref{isingrho}), (\ref{Stxexp}) and (\ref{SVNtxexp}) can be identified with the expansion of a $c=1/2$ theory. However, the parameter of this expansion, $\tx$, has meaning only within Baxter's parametrization of the model (\ref{eq:XYZ3bis}). To gain generality, the entropy is normally measured as a function of a {\it universal} parameter, such as the {\it correlation length} or the {\it mass gap}.

In the scaling limit, up to a multiplicative constant, one has:
\be
   \tx \approx \left( {\xi \over a_0} \right)^{-2} \approx \left( {\Delta E \over J} \right)^2 \; ,
   \label{xileading}
\ee
so that, substituting this into (\ref{Stxexp}), we get (\ref{Sxiexp}). Relation (\ref{xileading}) is crucial in turning the leading coefficient in the entropy of  a $c=1/2$ entropy such as (\ref{Stxexp}) into that of a $c=1$ theory, but it also doubles all the exponents of the subdominant corrections.
It is reasonable to assume that this $c=1$ model is some sort of double Ising, but its operator content does not seem to match any reasonable $c=1$ model, since only even exponent states exist.
Moreover, comparing (\ref{Sxiexp}) with (\ref{Sexp}), one would conclude that a $h=2$ operator is responsible for the first correction. It was argued in Ref. \cite{cardy2010} that a marginal field gives rise to logarithmic corrections in the entropy, thus, either this corrections is due to descendant of the {\it identity} (namely, the stress-energy tensor), or we should think of it as a $2h/\alpha$, with $h=1$.

In fact, we can write the partition function of the eight-vertex model as a bulk Ising model (i.e., quadratic in characters). Starting from (\ref{dualqprodZ}), we have
\bea
  \hskip -.5cm  {\cal Z} & = & {1 \over \sqrt{2}} x^{-{1 \over 12}} \xi^{{1 \over 12}}
  \prod_{k=1}^\infty \left( 1 - \xi^{1-2k} \right) \left( 1 + \xi^{1-2k} \right)
  \nonumber \\
  & = & {x^{-{1 \over 12}} \over \sqrt{2}} \left[ \chi^{\rm Ising}_{1,1} \left( {\ii \over \pi} \ln \xi \right)
  + \chi^{\rm Ising}_{2,1} \left( {\ii \over \pi} \ln \xi \right) \right]
  \nonumber \\
  && \qquad \times
  \left[ \bar{\chi}^{\rm Ising}_{1,1} \left( {\ii \over \pi} \ln \xi \right)
  - \bar{\chi}^{\rm Ising}_{2,1} \left( {\ii \over \pi} \ln \xi \right) \right]
  \label{bulkIsing} \\
  & = & {x^{-{1 \over 12}} \over \sqrt{2}} \left[ \left| \chi_{0}  \right|^2 - \left| \chi_{1/2} \right|^2
  - \chi_0 \bar{\chi}_{1/2} + \chi_{1/2} \bar{\chi}_0 \right] \; .
  \nonumber
\eea
This formulation provides a simple explanation of the Renyi entropy expansion (\ref{Sxiexp}) and its operator content. In fact,  it interprets the first correction as the Ising energy operator, and not as a descendant of the identity.\\
It also means that the prefactor in front of the logarithm in the entropy can be interpreted as ${c + \bar{c} \over 12}$ with $c = \bar{c} = {1 \over 2}$.

No fundamental reason is known for which CTM spectra (and partition functions) of integrable models can be written as characters in terms of the mass parameter $x$ or $\tilde{x}$. This is the case  also for Eq. (\ref{bulkIsing}).  Thus, so far, we can only bring forth this observation while any connection with some underlying Virasoro algebra remains to be discovered. To the contrary, sufficiently close to a critical point, the CTM construction can be seen as a boundary CFT and thus its character structure as function of the size of the system is dictated by the neighboring fix point. \cite{cardyle}

We are led to conclude that, in the scaling limit, the entropy can be written as a function of two variables: ${\cal S}_\alpha \left( {\ell \over a_0}, {\xi \over a_0} \right)$. When the inverse mass is larger that the subsystem size, we have the usual expansion of the form (\ref{Sncriticalexp}). But when the correlation length becomes the infrared cut-off scale, apparently a different expansion is possible, which, unlike (\ref{Sexp}), can contain terms with different exponents, like in (\ref{Sxiexp}). Hence, while the leading universal behavior has always the same numerical value and scales like the logarithm of the relevant infra-red scale, the exponents of the corrections might be in principle different for terms in $\ell$ and in $\xi$.

In the scaling limit, $a_0 \to 0$, $J \to \infty$, and $\tx \to 0$ in such a way to keep physical quantities finite. In this limit, only the scaling relation (\ref{xileading}) survives. However, at any finite lattice spacing $a_0$, there will be corrections which feed back into the entropy and that can be relevant for numerical simulations. To discuss these subleading terms we have to consider two regimes separately.

\subsection{Free Excitations: $0\leq \mu \leq \frac{\pi}{2}$}

For $0\leq \mu \leq \frac{\pi}{2}$ the lowest energy states are free, with dispersion relation (\ref{scaling7}).
To express the entropy as a function of the correlation length we need to invert (\ref{xibound}). This cannot be done in closed form. So we have to first expand (\ref{xibound}), finding
\bea
   {a_0 \over \xi } & = & 4 \tx^{1/2} \sum_{n=0}^\infty \sigma_{-1} (2n+1) \: \tx^{n}
   \nonumber \\
   & = & 4 \tx^{1/2} + {16 \over 3} \tx^{3/2} +  {24 \over 5} \tx^{5/2} + \ldots
   \label{xifree1}
\eea
and  then to invert this (by hand) to the desired order:
\be
   \tx = {1 \over 16} {a_0^2 \over \xi^2} \left[ 1 - {1 \over 6} \: {a_0^2 \over \xi^2}
   + {7 \over 144} \: {a_0^4 \over \xi^4} + \Ord \left( \xi^{-6} \right) \right] \: .
	\label{eq:inv}
\ee
We get:
\bea
   {\cal S}_\alpha & = & {1 + \alpha \over 12 \alpha} \ln {\xi \over a_0}
   + {1 - 2 \alpha \over 6 \alpha} \ln 2
   \nonumber \\
   && + B_\alpha \xi^{-{2 \over \alpha}} + C_\alpha \xi^{-2{ 1 + \alpha \over \alpha}}
   + B'_\alpha \xi^{-{4 \over \alpha}} + \ldots
   \nonumber \\
   && - \alpha B_\alpha \xi^{-2} - \alpha B'_\alpha \xi^{-4} + \ldots
   \label{Sfreeexp}
\eea
where the coefficients only depend on $\alpha$ and contain the proper power of $a_0$ to keep each term dimensionless [for instance $B_\alpha= {1 \over \alpha -1} \left( {a_0 \over 4} \right)^{2/\alpha}$].
We note that a new term has appeared (and more will be seen at higher orders) and that it is not of the forms discussed in Ref.\cite{camplostrini2010}.

It is worth noticing that if we express the mass-gap (\ref{freegap}) as function of $\tx$, we would get a different series expansion and thus different corrections to  the entropy. These subleading terms would have a different form, compared to (\ref{Sfreeexp}), and even be path dependent on how one approaches the critical point. We prefer not to dwell into these details now, postponing the description of this kind of path-depended behavior with the bound state's correlation length to the next section.

\subsection{Bound states: $\frac{\pi}{2} < \mu < \pi $}

For $\mu > \pi/2$, bound states become stable, and the lightest excitation becomes the $s=1$ state with dispersion relation (\ref{scaling11}). Accordingly, the expression for the correlation length is different in this region from before. Using the dual variable $\tx$ and the formulation of elliptic functions as infinite products, we can write (\ref{xibound}) as
\bea
   \hskip-2cm{a_0 \over \xi } & = & \ln { \left( - \tx^{1+ \tau \over 2}; \tx \right)_\infty \left( - \tx^{1- \tau \over 2}; \tx \right)_\infty \over
   \left( \tx^{1+ \tau \over 2}; \tx \right)_\infty \left( \tx^{1- \tau \over 2}; \tx \right)_\infty }
   \label{xibound1} \\
   & = & 4 \sum_{n=1}^\infty \sum_{k=1}^\infty {1 \over 2k -1} \cos \left[ {\pi^2 \over 2 \mu} (2k-1) \right] \tx^{(2n-1)(2k-1) \over 2} \:.
   \nonumber
\eea

\begin{figure}
   \begin{centering}
   \includegraphics[width=8cm]{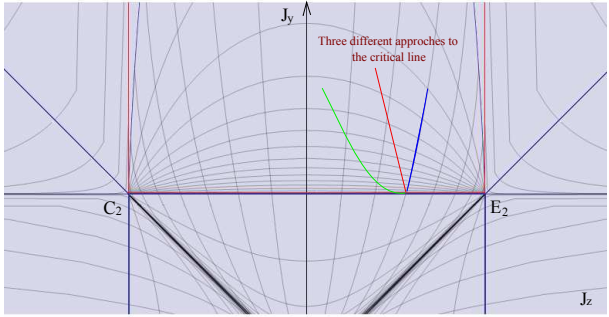}
   \end{centering}
   \caption{Three different ways to approach the critical line: along the $\mu$-constant lines (blue line), which is  referred to as renormalization group flow in the text; along straight lines (red line); along lines that approach the criticality with zero derivative (green line).}
   \label{fig:approching}
\end{figure}

The major difference in (\ref{xibound1}) compared to (\ref{xifree1}) is that the bound state correlation length does not depend on $\tx$ alone, but separately on $\mu$ and $\tau$. This means that in inverting (\ref{xibound1}) to find $\tx$ as a function of $\xi$, we have to first specify a relation between $\mu$ and $\tau$, i.e. to choose a path of approach to the critical line. We will follow three different paths, that are shown in  Fig.~\ref{fig:approching}.

\begin{enumerate}[(a)]

\item Renormalization group flow: The first natural path is (represented in the blue line in Fig.~\ref{fig:approching}):
\be
   \tau=\ii s \; ,  \qquad  \qquad
   \mu = \mu_0 \; ,
  \label{scaling12}
\ee
where $s \to 0$ guides our approach to the gapless point. This path, keeping $\mu$ fixed, corresponds to the RG flow. In the scaling limit, the XYZ chain is described by a sine-Gordon model, where $\mu$ is proportional to the compactification radius \cite{Luther}. Thus, assuming (\ref{scaling12}) means changing the bare mass scale, without touching $\beta$. In the $(J_z,J_y)$ plane, this path asymptotically crosses the critical line with slope $m = -2 / \cos \mu_0$.
Substituting this in (\ref{xibound1}) we get:
\be
  \label{scaling13}
   {a_0 \over \xi} =  4 g (\mu_0) \: \tx^{1/2} + {16 \over 3} g^3 (\mu_0) \: \tx^{3/2} +\Ord \left( \tx^{5/2} \right)  \; ,
\ee
where $ g(\mu) \equiv \cos {\pi^2 \over 2 \mu }$. Comparing (\ref{scaling13}) with the free case (\ref{xifree1}) we immediately conclude that the entropy retains an expansion similar to (\ref{Sfreeexp}), with the difference that all the coefficients now depend on $\mu_0$ and thus change along the critical line:
\bea
  \hskip-1cm{\cal S}_{\alpha} & \simeq & \frac{1 + \alpha}{12 \alpha} \ln \xi  + A_{\alpha} (\mu_0) + B_\alpha (\mu_0) \xi^{-{2 \over \alpha}}
  \nonumber \\
  && - \alpha B_\alpha (\mu_0) \xi^{-2} + C_\alpha (\mu_0) \xi^{-2 - { 2 \over \alpha}} + \ldots
  \label{scaling14}
\eea

\item  Straight lines in $(J_z,J_y)$ space: Let us now approach a conformal critical point exactly linearly in the $(J_z,J_y)$ plane:
\be
   \label{scaling24}
   J_y = 1+m \cdot s \; , \qquad \qquad
   J_z = s - \cos \mu_0  \; .
\ee
This path corresponds to the following parametrization of $\tau$ and $\mu$ (an example of which is the red line in Fig.~\ref{fig:approching}):
\be
   \label{scaling31}
   \begin{cases}
   \tau = -\ii \frac{\pi}{\ln(s)} + \Ord \left( \frac{1}{\ln^{2} s }\right) \; , \\
   \mu = \mu_0 + r(m,\mu_0) \cdot s + \Ord (s^{2}) \; ,
   \end{cases}
\ee
where $ r (m,\mu) \equiv {2 + m \cos \mu \over 2 \sin \mu }$.
Thus, in the limit $s \to 0$, the entropy parameter $\tx$ vanishes like $\tx \propto s^{\pi/\mu_0}$.
Using (\ref{scaling31}) in (\ref{xibound1})
\be
  \label{scaling34}
  {a_0 \over \xi } \simeq  4 g ( \mu_0 )  \tx^{1 \over 2}
  +  4 r (m,\mu_0)  g' (\mu_0) \tx^{\frac{1}{2}+\frac{\mu_0}{\pi}} + \ldots
\ee
Inverting this relation, we arrive at the following expansion of the R\'enyi
	 entropy along (\ref{scaling24})
\bea
   \label{scaling36}
   \hskip-2cm{\cal S}_{\alpha} & \simeq & \frac{1 + \alpha}{12 \alpha} \ln \xi + A_{\alpha} (\mu_0) \\
   & & + B_{\alpha} (\mu_0) \xi^{-2/\alpha}  + D_{\alpha} (m,\mu_0) \xi^{-2\mu_0/\pi} \ldots
   \nonumber
\eea
We notice that the last term yields a new type of correction, with a non-constant exponent, which varies with $\mu$. This term is of the form $\xi^{-(2-h)}$, where $h$ here is the scaling dimension of the vertex operator ${\rm e}^{\ii\beta\phi}$ of the underlying sine-Gordon theory. To the best of our knowledge, this is the first time that such a correction in the R\'enyi entropy of gapped systems is discussed and it also differs from those discussed in Ref. \cite{camplostrini2010}. It is surprising to see the appearance of the operator content of the underlying sine-Gordon field theory in the exponents of the expansion of $\mathcal{S}_{\alpha}$ on the lattice, whilst these operators do not enter in the scaling limit (\ref{Sxiexp}). We do not have a satisfactory understanding of this result, but we believe that an approach similar to that of Ref.\cite{cardy2010} might clarify the point.

Since $\frac{\pi}{2}< \mu_0 < \pi$, the exponent of this new correction ranges between $1$ and $2$ and always dominates over the $\xi^{-2}$ term in (\ref{scaling14}) and competes with the correction $\xi^{-2/\alpha}$ for $\alpha < 2$.
Notice, however, that for $m_0 = -2 / \cos \mu_0 $, $r (m_0,\mu_0) =0$ and thus $D_\alpha (m_0,\mu_0) =0$: this new correction disappears. This is precisely the slope that corresponds to the RG flow we considered before. Also, $g'(\pi/2) = 0$ and thus the coefficient in front of this new correction vanishes at the crossing point between the free and the bound state. Thus, we can conclude that this new term is turned on by the existence of a bound state and it is a clear signature of a truly interacting theory. Moreover, the path one chooses to approach criticality selects a scaling limit in which irrelevant operators can be generated and these can modify the perturbative series that defines the correlation length, leading to something like we observed.

\item  Straight lines in $(l,\mu)$ space: Finally, as a generalization of the first case, let us consider the straight line
\be
   \label{scaling18}
   \tau=\ii s \: , \qquad \qquad
   \mu = \mu_0 + r \cdot s \: ,
\ee
where $r$ is the slope in the $(\tau,\mu)$ plane. Interestingly, this trajectory maps into a curve in the $(J_z,J_y)$-plane approaching the critical point with zero derivative, i.e. with a purely quadratic relation in a small enough neighborhood of the conformal point (see for instance the green line in Fig.~\ref{fig:approching}).
Putting (\ref{scaling18}) in (\ref{xibound1}) we have
\bea
   \hskip -1.5cm{a_0 \over \xi } & \hskip-0.2cm= & \hskip-0.2cm 4 g (\mu_0) \tx^{1/2}
   +  4 r {\textstyle {\pi^2 \over \mu_0 }}  g' ( \mu_0)  {\tx^{1/2} \over \ln \tx}
   \nonumber \\
   && + {16 \over 3} g^3 ( \mu_0) \tx^{3/2} {\textstyle + \Ord \left( {\tx^{1/2} \over \ln^2 \tx},  {\tx^{3/2} \over \ln \tx} \right)  } \; .
   \label{scaling21}
\eea
We notice the appearance of a strange logarithmic correction in the expansion. Inverting (\ref{scaling21}) and plugging it into the entropy we get
\be
  \label{scaling23}
  {\cal S}_{\alpha} = \frac{1 + \alpha}{12 \alpha} \ln \xi  + A_\alpha (\mu) +\frac{E_{\alpha}(r,u)}{\ln \xi } + \ldots \: ,
\ee
This kind of logarithmic corrections were found also in Ref. \cite{calabrese2010}, but only in studying the limit $\alpha \to\infty$, i.e. the so-called single-copy entropy. They were also predicted in Ref. \cite{camplostrini2010} as signatures of marginal operators in a CFT, but with a different power. From our results, it seems that this kind of very unusual corrections may appear at finite $\alpha$'s, by selecting a proper path  approaching the critical point. This can be due to the presence of a marginally-relevant operator in the theory, generated when reaching the critical line through a zero-slope curve and corresponding to a changing of the compactification radius. We will attempt a field theoretical analysis of this kind of term in a future work.
\end{enumerate}

\section{Conclusions and outlooks}
\label{sec:end}

By using the example of the integrable {\em XYZ} chain, we proved that, for a massive model, the study of the corrections to the entanglement entropy as a function of the correlation length requires a separate analysis from the one that yields the entropy as function of the subsystem size.

For the bipartite R\'enyi entropy of the {\em XYZ} model of a semi-infinite half-line we found, in the scaling limit and as a function of the correlation length, the universal form (\ref{Sxiexp}), where all subleading contributions are explicitly written, thanks to a novel formulation of the reduced density matrix in terms of {\em q}-products. We argued that these corrections are best interpreted in light of a previously unnoticed bulk Ising structure of the CTM formulation of the model. This means that corrections as a function of the correlation length have different exponents compared to those depending on the length of the subsystem, unlike what was expected from previous studies. This also implies that the coefficient ${c + \bar{c} \over 12}$ of the logarithmic leading term has the same value both for $c = \bar{c} = 1/2$ of the bulk Ising formulation in the mass parameter and for the $c=1$, $\bar{c}=0$ of the critical chiral free boson model in the subsystem size.

In this respect, it is also interesting to note that the reduced density matrix $\hat{\rho}$ of (\ref{XYZrho}) can be written as \cite{Peschel1, Peschel2, Peschel3, Peschel4}:
\be
\label{thermal}
\hat{\rho}\propto {\rm e}^{-H_{{\rm CTM}}}, \qquad H_{{\rm CTM}}=\sum_{j=1}^{\infty}2\epsilon_{j}\eta_{j}^{\dag}\eta_{j}
\ee
where $(\eta_{j}^{\dag},\eta_{j})$ are (Majorana) fermionic creation and annihilation operators for single particle states with eigenvalue $2\epsilon_{j}=2j\epsilon$ (note that $H_{{\rm CTM}}$ is not the Hamiltonian of the subsystem $A$).
This representation strongly supports the interpretation that the $c=1$ theory is constructed in terms of $c=1/2$ (Majorana) characters.

In Refs.\cite{camplostrini2010, alcaraz2011} it was shown that the first correction in the R\'enyi entropy of a critical {\em XXZ} chain as a function of the subsystem size $\ell$ goes like $\ell^{-2 K / \alpha}$, where $K$ is the Luttinger parameter of the model. This fact has also been checked in many other critical $c=1$ quantum spin chain models via DMRG simulations\cite{fagotti,alcaraz2011,dalmonte2011}.
	When going to  the corresponding massive model, assuming that the operators responsible for the corrections remain the same, the simple scaling prescription \cite{calabrese2010} would give a term of the type $\xi^{-K/\alpha}$. The results presented here would then indicate an improbable fixed value for the Luttinger parameter $K=2$. This exponent for the massive {\em XXZ} chain was observed before (take, for instance, Ref. \cite{calabrese2010}), but its nature has not been discussed.
	Instead, consistently with Ref. \cite{alcaraz2011}, we found in Sec. \ref{se:expansion} that the operator responsible for this correction is the energy of the underlying bulk Ising model.
	As pointed out in Ref.\cite{cardy2010}, the leading correction in $\ell$ is of the form $\ell^{-2K/\alpha}$ in the one-interval case and $\ell^{-K/\alpha}$ for the half-line, whereas for two intervals, in the Ising case, the exponent acquires an additional factor of $2$, which counts the number of twist fields at the edge of the interval \cite{fagotticala}. It would be interesting to perform a calculation for one interval with two boundary points in our case too, to check whether a doubling of the exponent in the correlation length would happen in this case as well.
	We also observe that our results stem from the study of the formally conformal  structure emerging in the R\'enyi entropy of the ground state. It would be of great interest to perform a similar analysis for some excited states.

We also showed that, if one takes into account lattice effects, there is a proliferation of new standard and {\em unusual} corrections, which in general are path-dependent, and can assume the forms: $\xi^{-2h}$, $\xi^{-2h/\alpha}$, $\xi^{-(2-h)}$, $\xi^{-2(1+1/\alpha)}$, or even $1/\ln(\xi)$, where $h$ is the scaling dimension of a relevant operator of the critical bulk theory.
We conjecture that the last logarithmic correction is a consequence of some marginally-relevant operator in the theory.
In addition each of the previous terms appears multiplied by a point-dependent and $\alpha$-dependent coefficient that can vanish in even large parametric regions, preventing the corresponding unusual correction from showing up in the scaling limit of the R\'enyi entropy.
We should remark that the same analysis, carried out in terms of the mass-gap instead of the correlation length, would generate even different corrections. These differences will be analyzed elsewhere for the {\em XYZ} chain, but are a general feature of lattice models that have to be taken into account in numerical analysis.

In Ref. \cite{camplostrini2010} it was given a list of possible subleading contributions to the R\'enyi entropy for a CFT. From a na\"ive scaling argument, one could expect the same kind of terms to appear in a massive theory sufficiently close to criticality. However, some of the corrections we observed do not fit this expectation. This could be due to strictly ultraviolet effects that cannot be captured by a QFT or to a need to improve the scaling argument. We believe that an analysis similar to that carried out for critical systems in Ref. \cite{cardy2010} could be successfully applied to a massive sine-Gordon theory, by introducing an ultra-violet cut-off (the lattice spacing) to regularize divergent integrals and extract the corrections from the counterterms\footnote{An interesting alternative field theoretical approach to massive models has been presented in: \\
	J.L. Cardy, O.A. Castro-Alvaredo, and B. Doyon, J. Stat. Phys. \textbf{130} (2007) 129,  O.A. Castro-Alvaredo, and B. Doyon, J. Phys. A \textbf{41}  (2008) 275203, B. Doyon, Phys. Rev. Lett. \textbf{102} (2009) 031602.} .
This type of check would be dual to  what we have presented here: while in our approach we started from a lattice model to infer its universal behavior, in the other, one would start from a field theory to understand the origin of the correction. In conclusion, we believe that further analysis are needed to provide a consistent field theoretical interpretation of the corrections arising in massive models, especially for the relevancy of such problem in numerical studies.

\section*{Acknowledgments}

We wish to thank Fabian Essler, Andrea De Luca, Pasquale Calabrese, John Cardy, Maurizio Fagotti, Emanuele Levi, Ingo Peschel, Cristian Degli Esposti Boschi, Fabio Ortolani, Luca Taddia and Marcello Dalmonte for useful and very pleasant discussions. This work was partially supported by two INFN COM4 grants (FI11 and NA41) and by the U.S. Department of Energy under cooperative research Contract Number DE-FG02-05ER41360. FF was supported in part by a Marie Curie International Outgoing Fellowship within the 7th European Community Framework Programme (FP7/2007-2013) under the grant PIOF-PHY-276093.


\begin{thebibliography}{99}

\bibitem{Fazio}
 L. Amico, R. Fazio, A. Osterloch and V. Vedral, Rev. Mod. Phys. \textbf{80}, 517 (2008)

\bibitem{eisert}
  J. Eisert, M. Cramer, and M. B. Plenio, Rev. Mod. Phys. {\bf 82}, 277 (2010)

\bibitem{ccdrev}
 P. Calabrese, J. Cardy, and B. Doyon, J. Phys. {\bf A 42}, 500301 (2009)

\bibitem{Schumacher}
 C. H. Bennet, H. J. Bernstein, S. Popescu, B. Schumacher, Phys. Rev. A \textbf{53}, 2046 (1996)

\bibitem{nielsen}
 M. A. Nielsen, and I. L. Chuang; {\it Quantum Computation and Quantum Information}, Cambridge University Press
(2000)


\bibitem{Renyi}
  A. R\'enyi
	, {\em Probability Theory}, North-Holland, Amsterdam, 1970.

\bibitem{calabreselefevre2008}
  P. Calabrese, and A. Lefevre, Phys. Rev. {\bf A 78}, 032329 (2008)

\bibitem{franchini2010}
  F. Franchini, A. R. Its, V. E. Korepin, and L. A. Takhtajan, Quantum Inf. Processing {\bf 10}, 325 (2011)

\bibitem{Holzey}
  C. Holzhey, F. Larsen, F. Wilczek Nucl. Phys. B \textbf{424}, 443 (1994)

\bibitem{Cardy}
  P. Calabrese, and J. L. Cardy, J. Stat. Mech. \textbf{P06002} (2004)

\bibitem{calabrese2010}
  P. Calabrese, J. Cardy, and I. Peschel, J. Stat. Mech. 1009:P09003 (2010)

\bibitem{camplostrini2010}
  P. Calabrese, M. Campostrini, F. Essler, and B. Nienhuis, Phys. Rev. Lett. {\bf 104}, 095701 (2010)

\bibitem{cardy2010}
  J. Cardy, and P. Calabrese, J. Stat. Mech. P04023 (2010)

\bibitem{essler2010}
  P. Calabrese, and F.H.L. Essler, J. Stat. Mech. P08029 (2010)

 \bibitem{fagotti}
  M.F. Fagotti, and P. Calabrese, J. Stat. Mech. P01017 (2011)
	
\bibitem{alcaraz2011}
 J. C. Xavier, F. C. Alcaraz, Phys. Rev. {\bf B 83}, 214425 (2011); Phys. Rev. {\bf B 85}, 024418 (2012).

\bibitem{dalmonte2011}
 M. Dalmonte, E. Ercolessi, L. Taddia, Phys. Rev.  \textbf{B 84}, 085110 (2011) and arXiv:1105.3101


\bibitem{Sutherland}B. Sutherland, J. Math. Phys. \textbf{11}, 3183 (1970)
	

\bibitem{Baxter}
  R. J. Baxter, \emph{Exactly solved models in statistical mechanics}, Academic Press, London (1982)

\bibitem{McCoy}
  J.D. Johnson, S. Krinsky, and B.M. McCoy, Phys. Rev. \textbf{A 8}, 5 (1973)
	
\bibitem{comment} In Ref. \cite{McCoy} the correlation length close to the critical line is calculated for periodic boundary conditions as that of the $s = 1$ bound state in the disordered phase of the 8-vertex model. In the ordered phase this state is forbidden by superselection rules and one has to use the $s = 2$ bound state (both of which become breathers in the scaling limit). The CTM construction implies fixed boundary conditions and thus we can use the $s = 1$ breather, as the lightest excitation in the attractive regime of the sine-Gordon line of the {\em XYZ} model.
	
\bibitem{Evangelisti}
  E. Ercolessi, S. Evangelisti, and F. Ravanini, Phys. Lett. {\bf A 374}, 2101 (2010)  and arXiv:0905.4000

\bibitem{Ercolessi2011}
  E. Ercolessi, S. Evangelisti, F. Franchini, and F. Ravanini, Phys. Rev. {\bf B 83}, 012402 (2011) 	and arXiv:1008.3892
	
\bibitem{Peschel1}I. Peschel, M. Kaulke and O. Legeza, Ann. Phys.
(Leipzig) \textbf{8} 153 (1999)

\bibitem{Nishino1}T. Nishino, J. Phys. Soc. Japan \textbf{64}, 3598 (1995)


\bibitem{Nishino2}T. Nishino and K. Okunishi, J. Phys. Soc. Japan
\textbf{66}, 3040 (1997) 	
	
\bibitem{Wittaker} E. T. Whittaker and G. N. Watson, {\em A Course of Modern Analysis},
Cambridge at the University Press 1927

\bibitem{Luther}A. Luther, Phys. Rev. \textbf{B14}, 2153 (1976)
	
\bibitem{Franchini}
 F. Franchini, A. R. Its and V. E. Korepin, J. Phys. A: Math. Theor. \textbf{41}, 025302 (2008)

\bibitem{Franchini2}
 F. Franchini, A. R. Its, B-Q. Jin and V. E. Korepin,J. Phys. A: Math. Theor. \textbf{40}, 408467 (2007)

\bibitem{Saleur} H. Saleur and M. Bauer, Nuclear Physics {\bf B 320}, 3 (1989)	

\bibitem{cardyle}
  J.L. Cardy, Les Houches Summer School Lectures (1988)
	
\bibitem{apostolbook}
   T.M. Apostol, \emph{Introduction to analytic number theory}, Springer-Verlag, New York-Heidelberg (1976)
	
\bibitem{Affleck}J. Affleck and A. W. W. Ludwig, Phys. Rev. Lett.
\textbf{67}, 161 (1991)

\bibitem{Peschel2}I. Peschel, arXiv:1109.0159

\bibitem{Peschel3}I. Peschel and V. Eisler, J. Phys. A: Math. and Theory. \textbf{42}, 504003(2009)

\bibitem{Peschel4}M.C. Chung, I. Peschel, Phys. Rev. \textbf{B 64}, 064412 (2001)

\bibitem{fagotticala}
  M. Fagotti, P. Calabrese, J. Stat. Mech. {\bf 2010}, P04016 (2010) and arXiv:1003.1110







\end{thebibliography}
\end{document}